# Target echo strength modelling at FOI, including results from the BeTSSi II workshop

Martin Östberg[1]

[1] Swedish Defense Research Agency
Contact email: martin.ostberg@foi.se

## Abstract

An overview of the target echo strength (TS) modelling capacity at the Swedish Defense Research Agency (FOI) is presented. The modelling methods described range from approximate ones, such as raytracing and Kirchhoff approximation codes, to high accuracy full field codes including boundary integral equation methods and finite elements methods. Illustrations of the applicability of the codes are given for a few simple cases tackled during the BeTTSi II (Benchmark Target Echo Strength Simulation) workshop held in Kiel 2014.

## 1 Introduction

The target echo strength (TS) is an established measure of a submarine's (or any submerged object's) reflection of an incoming acoustic wave. It is widely used when assessing a submarine's ability to avoid detection.

A natural way of investigating and determining TS is by measurements, which are often costly and cumbersome. As a cost effective complement to measurements, numerical modelling is an attractive way of evaluating and improving the TS.

In order to address this, FOI has developed a range of codes designed to predict the TS of underwater objects. These codes range from approximate ones, such as raytracing and Kirchhoff approximation codes, to high accuracy full field codes including boundary integral equation methods and finite element methods.

In this paper, a brief description of these codes and the underlying theories are given. As illustrations of the capabilities of the codes a few results from the BeTSSi II (Benchmark Target Echo Strength Simulation) workshop held in Kiel in 2014 are presented. The workshop was initiated by FWG, Reasearch Department for Underwater and Marine Geophysics, Kiel, as an effort to jointly evaluate the simulation capacity at hand by a number of organisations.

## 2 Target echo strength

Given a plane wave impinging on a scattering object at an in-plane angle $\varphi_1$ and an out-of-plane angle $\theta_1$, according to Figure 1, the target echo strength is given in terms of the scattered pressure $p_{\text{sc}}(r, \varphi_2, \theta_2)$ a distance $r$ from the scatterer relative the impinging wave

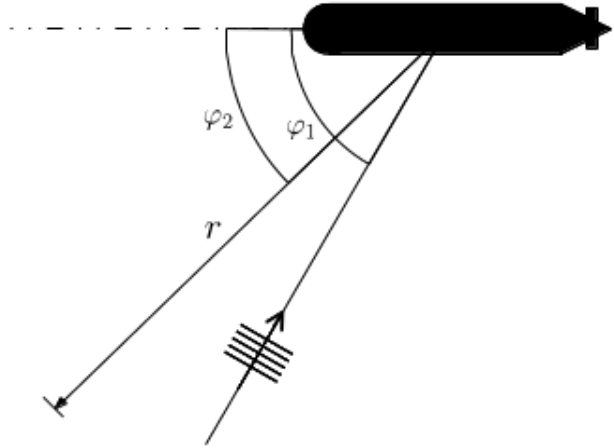

Figure 1: Geometric definitions for TS in the horizontal plane, *i.e.* $\theta_1 = \theta_2 = 0$.

pressure, $p_{\text{in}}$, as

$$\text{TS}(\varphi_1, \theta_1, \varphi_2, \theta_2) = \lim_{r \to \infty} 20\log_{10}\left(\frac{r|p_{\text{sc}}|}{r_0|p_{\text{in}}|}\right) \tag{1}$$

where $r_0 = 1$ m is a reference radius. Furthermore, the *monostatic* TS is obtained when

$$\varphi_1 = \varphi_2 \tag{2}$$

and

$$\theta_1 = \theta_2 \tag{3}$$

while

$$\varphi_1 \neq \varphi_2 \tag{4}$$

or

$$\theta_1 \neq \theta_2 \tag{5}$$

gives the *bistatic* TS.

## 3 Modelling approaches

When computing the scattered field from a submerged body, a number of methods are at hand. They differ in accuracy, valid frequency range and in the required computational resources. Below, three classes of standard methods are distinguished: full field methods, methods based on the Kirchhoff scattering approximation and raytracing methods.

### 3.1 Full field methods

The term full field methods describes a group of methods which solve the wave equation without introducing approximations. Two main methods can be distinguished: boundary integral equation (BIE) methods and volume finite element methods (FEM), where the former is the most frequently used at FOI. The boundary integral equation methods rely on the external Helmholtz-Kirchhoff integral representation [1] of the wave field outside a closed surface $\Gamma$,

$$\int_\Gamma (G(\boldsymbol{r},\boldsymbol{r}_0)\nabla p(\boldsymbol{r}) - p(\boldsymbol{r})\nabla_1 G(\boldsymbol{r},\boldsymbol{r}_0)) \cdot \boldsymbol{n} \mathrm{d}\Gamma(\boldsymbol{r}) = p(\boldsymbol{r}_0) - p_{\mathrm{in}}(\boldsymbol{r}_0), \tag{6}$$

relating the total pressure, $p(\boldsymbol{r}_0)$, at an arbitrary field point $\boldsymbol{r}_0$ outside the scatterer to the pressure at the points $\boldsymbol{r}$ on the surface of the scatterer. Letting the field point approach the surface $\boldsymbol{r}_0 \to \boldsymbol{r}_\mathrm{s}$, an equation for the unknown surface pressure is obtained,

$$\int_\Gamma (G(\boldsymbol{r},\boldsymbol{r}_\mathrm{s})\nabla p(\boldsymbol{r}) - p(\boldsymbol{r})\nabla_1 G(\boldsymbol{r},\boldsymbol{r}_\mathrm{s})) \cdot \boldsymbol{n} \mathrm{d}\Gamma(\boldsymbol{r}) = \frac{1}{2}p(\boldsymbol{r}_\mathrm{s}) - p_{\mathrm{in}}(\boldsymbol{r}_\mathrm{s}). \tag{7}$$

Given the incident pressure field, $p_{\mathrm{in}}$, and the free space Green's function, $G(\boldsymbol{r},\boldsymbol{r}_0)$, the integral in Eq. (7) can be discretized and solved to yield the sought surface pressure $p(\boldsymbol{r}_\mathrm{s})$, which is inserted in Eq. (6) to obtain the far field solution.

At FOI, two codes falling into the category of full field methods are considered. The first code, XFEM_BIE, is a boundary integral code using a collocation method with high order B-spline basis functions [2]. It uses a smooth representation of the scatterer. Although applicable to scatterers with some internal structure, *e.g.* thin elastic shells, alternative methods are often more suitable to tackle this problem. One way, is to combine XFEM_BIE with the commercial finite element software Comsol Multiphysics into a code named BIE_FEM. By using volume finite elements for the inner problem and the high order B-spline representation of the field on the surface $\Gamma$, an arbitrary level of complexity of the scatterer is in principle allowed.

## 3.2 Kirchhoff methods

A way of reducing the computational load resulting when employing full field methods is by making an assumption on the surface pressure resulting from a given incident pressure field. A popular approach is to use the so called Kirchhoff-approximation. Splitting the pressure field at the scatterer in the incoming and scattered field, $p = p_{\mathrm{in}} + p_{\mathrm{sc}}$, a plane wave reflection coefficient for the illuminated side of the scatterer, $R$, is assumed as

$$Rp_{\mathrm{in}} = p_{\mathrm{sc}} \tag{8}$$

and

$$R\nabla p_{\mathrm{in}} \cdot \boldsymbol{n} = -\nabla p_{\mathrm{sc}} \cdot \boldsymbol{n}, \tag{9}$$

while $p = 0$ on the non-illuminated side of the scatterer. Two different codes based on the Kirchhoff-approximation are here presented: ARTES [3] and XFEM_KIRTRI, the latter a FOI-code written by Ilkka Karasalo. The codes are based on a similar approach; the scatterer is represented by triangular facets, $S$, in which the integral in Eq. (6) is approximated by

$$\int_S (a + b\eta + c\xi)\mathrm{e}^{\mathrm{i}(k_\eta \eta + k_\xi \xi)} \mathrm{d}S(\eta,\xi), \tag{10}$$

where $a$, $b$, $c$ can be obtained by evaluating the incoming pressure field in three points on the triangle and $k_\eta$ and $k_\xi$ are projections of the incoming field wavenumber on the local triangle coordinates $\eta$ and $\xi$. The integral (10) can then be computed exactly for each triangular facet and the total scattered pressure is obtained by summing the contributions from each facet. The formulation (10) has the advantage that the triangular facets can be chosen arbitrarily large for flat surfaces, provided that the source–scatterer and scatterer–receiver distances are much larger than the facet side lengths. The accuracy when computing the integral (6) with assumptions (8) and (9) for TS evaluations thus only depends on how well the curvature of the scatterer surface is represented by the facets.

### 3.3 Raytracing

Raytracing, sometimes referred to as geometrical optics, is a high frequency approximation of the acoustic field. The scattered field then consists solely of the specular reflection and its amplitude is determined by the radii of curvature at the point of reflection. A challenge when constructing a raytracing code is to obtain a smooth enough representation of the scattering object since a small local error in radii of curvature drastically can alter the TS. For example, triangular facets, as described in the previous section, is a poor choice, since they fail in representing the local radii of curvature. The code XRAY fixes this issue by mapping the unit sphere onto the scatterer and using high order B-splines. This smooth mapping minimizes the risk of spurious reflections that can result when using lower order surface representations (such as the previously mentioned piece-wise linear facets). On the other hand, it is not ideal for representing surfaces with vanishing radii of curvature, such as sharp corners.

## 4 Results and discussion

A few results from the BeTTSi II workshop are presented in the form of plots, depicting TS as function of receiver angle, $\varphi_2$, for the objects shown in Figures 2 and 3 [4]. Throughout, TS in the horizontal plane is considered, *i.e.* $\theta_1 = \theta_2 = 0$.

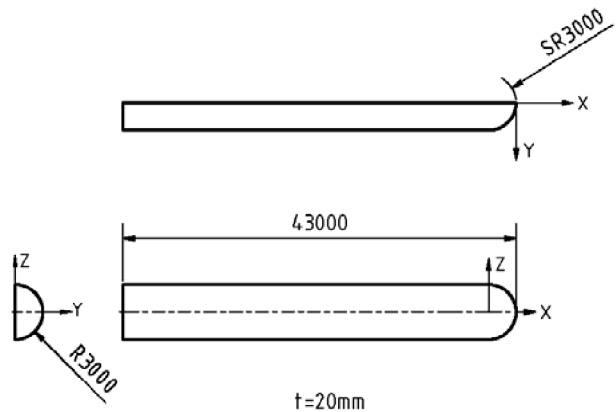

Figure 2: Dimensions of Model 1. All lengths are given in mm. The spherical coordinates used in the TS definition relate to the Cartesian coordinates $(X, Y, Z)$ as $X = r \cos \varphi \cos \theta$, $Y = r \sin \varphi \cos \theta$ and $Z = r \cos \theta$. Illustration taken from Ref. [4].

In Figure 4 the bistatic TS for model 1 assuming hard wall boundary condition (*i.e.*, assuming $R = 1$ in Eqs. (8) and (9)) is shown for two angles of incidence: $\varphi_1 = 240°$ and $\varphi_1 = 300°$. As reference solution, results from simulations with BIE_FEM are given and compared with approximate solutions obtained using XFEM_KIRTRI. As expected, the accuracy of the Kirchhoff approximation increases with frequency. At 500 Hz, both the specular reflection, the back lobe as well as most parts of the side lobes are captured with good confidence.

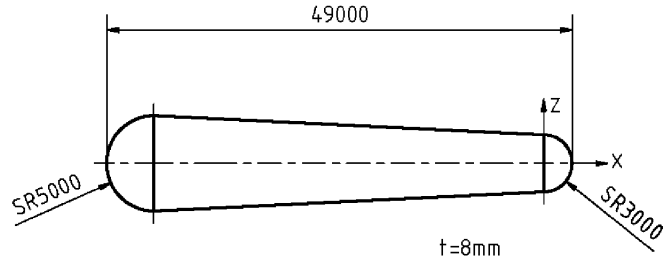

Figure 3: Dimensions of Model 3. All lengths are given in mm. See Figure 2 for angle definitions. Illustration taken from Ref. [4].

Similar observations are made for model 3 (Figure 5), with the additional observation that for this model, Kirchhoff approximation is more accurate than for model 1. This is probably due to the lack of sharp corners/edges in contrast to model 1. In the plots, results using both available full field codes, XFEM_BIE and BIE_FEM, are included. The difference in TS between these is negligible, verifying their correctness.

Simulations of the monostatic TS for model 3 are given in figure 6. Except for the broadside peaks at $\sim 87°$ and $\sim 183°$, the TS is dominated by the half-spherical end caps, and agrees well with the high frequency limit of a sphere ($20\log_{10}(a/2)$; $a$ being the sphere radius). The TS computed with XRAY is frequency independent and corresponding to the high frequency limit. The anomalies at $0°$ and $180°$, where completely flat curves are expected, are probably due to difficulties in obtaining a sufficiently smooth representation of the body, which is critical for correctly modeling high frequency scattering.

## 5 Conclusions

The results presented in this paper highlight some features of common modelling methodologies used when assessing the target strength of underwater objects. For general convex, rigid structures, the TS can be computed with good reliability for a broad frequency range. Full field codes provide accurate predictions for low to medium frequencies. For higher frequencies, where the increasing computational resources required often make full field methods unfeasible, raytracing and Kirchhoff methods can be used with reasonable confidence.

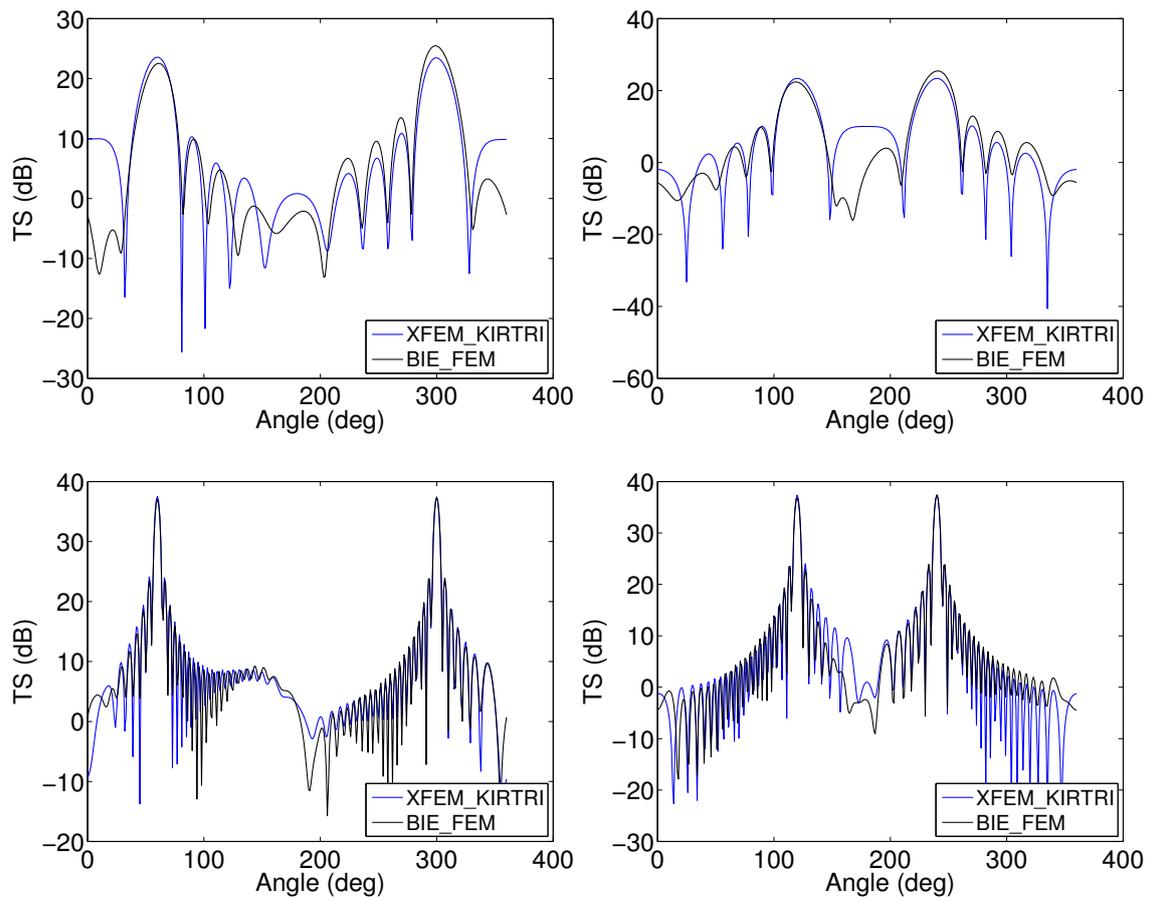

Figure 4: Bistatic TS for Model 1; $\varphi_1 = 240°$ (left column) and $\varphi_1 = 300°$ (right column) for 100 Hz (top) and 500 Hz (bottom).

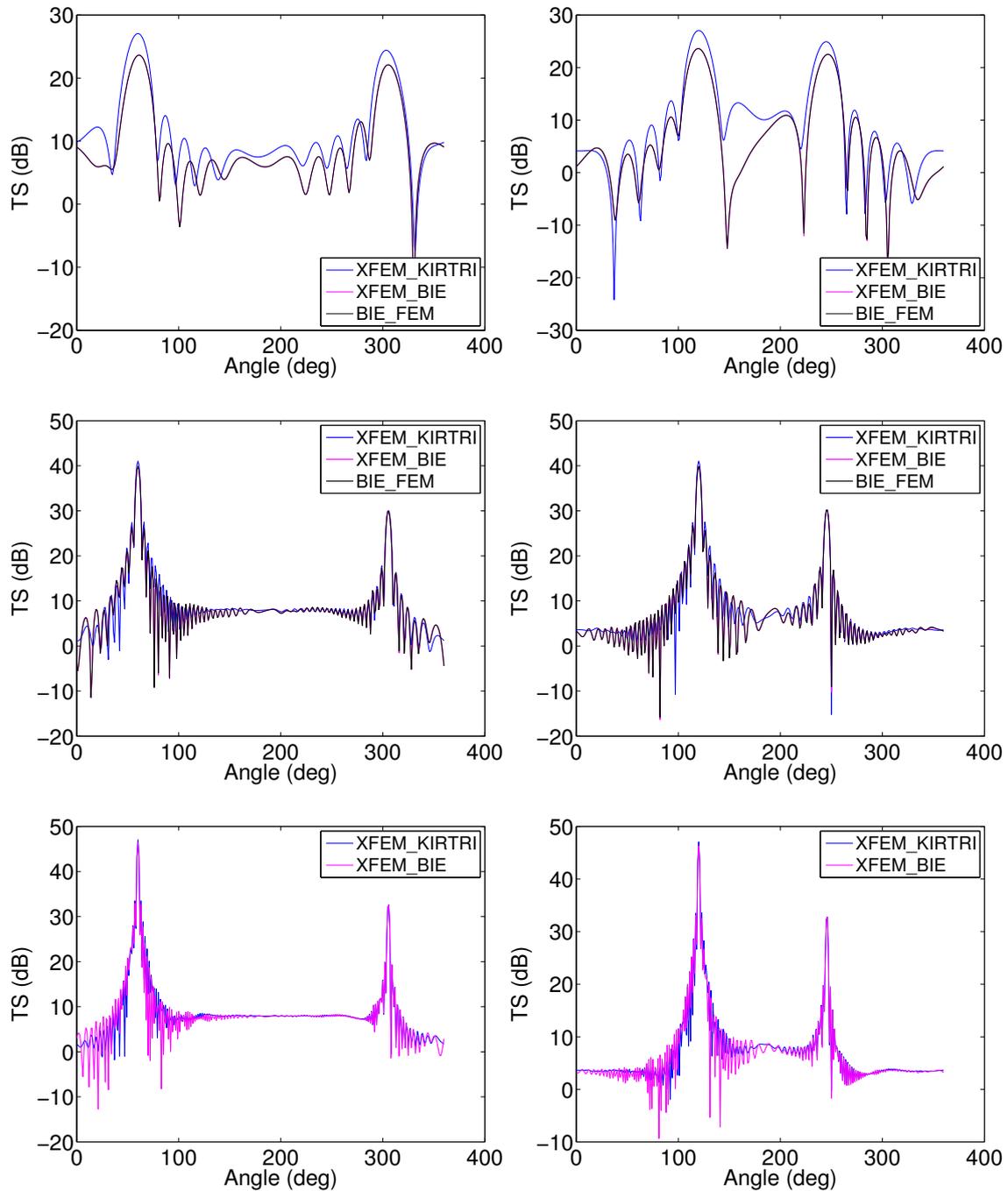

Figure 5: Bistatic TS for Model 3; $\varphi_1 = 240°$ (left column) and $\varphi_1 = 300°$ (right column) for 100 Hz (top), 500 Hz (middle) and 1 kHz (bottom).

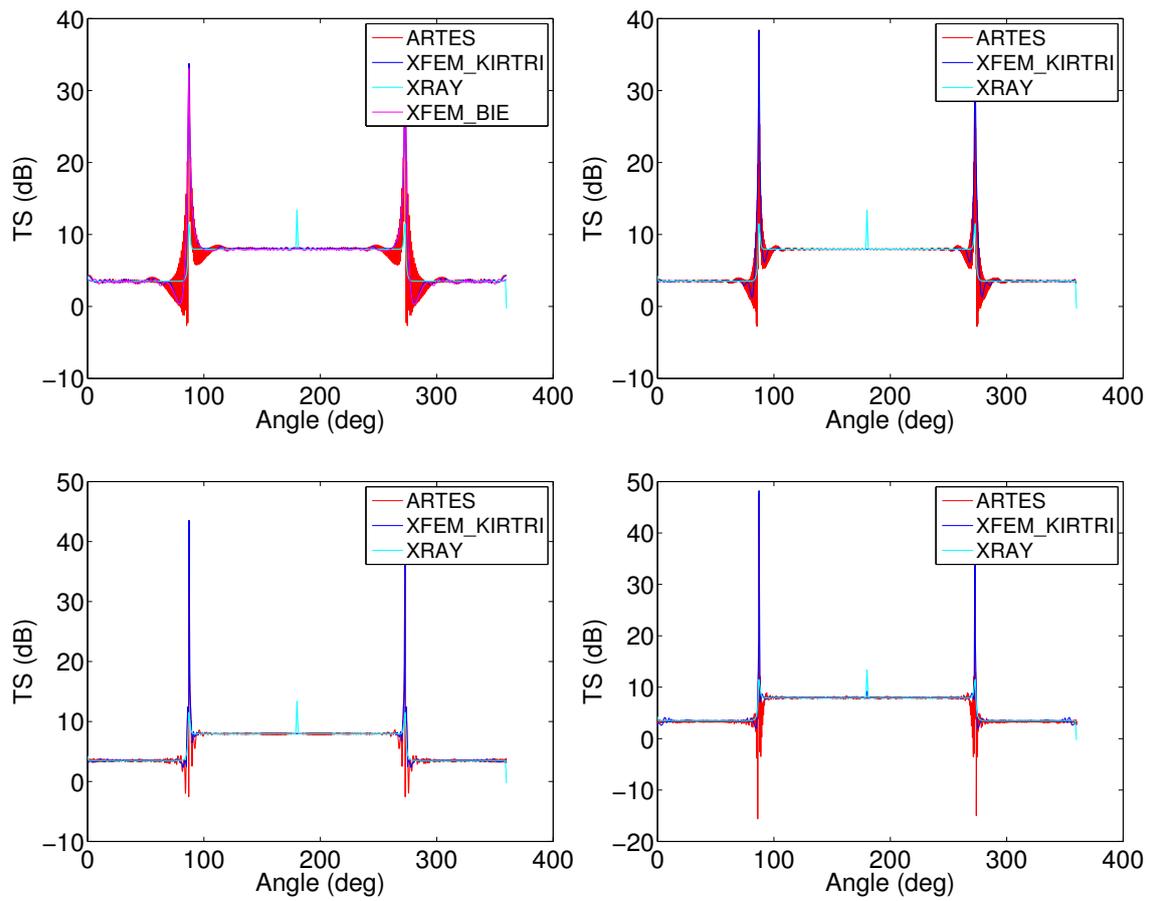

Figure 6: Monostatic TS for Model 3; 1 kHz (top left), 3 kHz (top right) and 10 kHz (bottom left) and 30 kHz (bottom right).